\documentstyle[preprint,tighten,aps,axodraw,float]{revtex}
\newcommand{\be}{\begin{equation}}
\newcommand{\ee}{\end{equation}}
\newcommand{\ba}{\begin{eqnarray}}
\newcommand{\ea}{\end{eqnarray}}

\newcommand{\fr}[2]{\frac{#1}{#2}}

\def\vec#1{{\mbox{\boldmath$#1$}}}

\newcommand{\k}{\mbox{$\vec{k}$}}

\newcommand{\r}{\mbox{$\vec{r}$}}

\newcommand{\lb}{\left (}
\newcommand{\rb}{\right )}

\newcommand{\ep}{\epsilon}

\begin{document}

\title{
%
%preprint number:
%
\[ \vspace{-2cm} \]
\noindent\hfill\hbox{\rm SLAC-PUB-8756 } \vskip 1pt
\noindent\hfill\hbox{\rm hep-ph/0101228} \vskip 10pt
%
% now the title:
%
%\title{
Radiative corrections to the Casimir force
and effective field theories
}

\author{Kirill Melnikov\thanks{
e-mail:  melnikov@slac.stanford.edu}}
\address{Stanford Linear Accelerator Center\\
Stanford University, Stanford, CA 94309}
\maketitle

\begin{abstract}
Radiative corrections to the Casimir force between two parallel plates
are considered in both  scalar field theory of one massless and one 
massive field and in QED.  Full calculations are contrasted with 
calculations based on employing ``boundary-free'' 
effective field theories. The difference between 
two previous results  on QED radiative corrections  
to the Casimir force between two parallel plates is clarified and 
the low-energy effective field theory for the Casimir effect 
in QED is constructed. 
\end{abstract}


\section{Introduction} 

Attraction of two infinitely conducting plates caused by vacuum 
fluctuations was predicted by H.~Casimir in 1948 \cite{casimir}. 
Experimental  attempts to verify this prediction have started almost 
immediately, however, it took almost half a century to reach  
conclusive and accurate outcome. To date the Casimir effect 
is verified on a few per cent level \cite{lamoroux,mohideen}.  

Recent theoretical work on the Casimir effect has been focused 
around several  topics \cite{milton} such as calculation of the Casimir force 
for complicated geometries, material dependence and 
the study  of radiative corrections to the Casimir force in QED. 
The latter studies resulted in an interesting controversy. 
The one-loop QED radiative corrections to the Casimir force in  case 
of two infinitely conducting parallel plates have been 
computed in \cite{bordag} and have been found to be ${\cal O}(\alpha/(mL))$
relative to the leading order result ($m$ is the electron mass,  
$L$ is the distance between the two plates and $\alpha$ is the fine structure 
constant). Later on,  the problem was reconsidered in \cite{kwong}. 
The authors of Ref.\cite{kwong}  argued that, 
since the Casimir force is the long-distance effect 
with a typical photon momentum being of the order $\sim 1/L$, 
it is natural to integrate out the momentum scales comparable 
to the electron mass and to use the effective field theory
of QED (in essence, the Euler-Heisenberg Lagrangian) 
to compute radiative corrections to the Casimir force. 
That calculation resulted in the relative correction 
${\cal O}(\alpha^2/(mL)^4)$ that  is  higher order in $\alpha$,
analytic in the mass of the electron squared and much more strongly 
suppressed by the distance between the two plates than the 
results of \cite{bordag}. 

To the best of my knowledge, the issue has never been clarified 
(see e.g. the recent discussion in \cite{ravndall}) and it  is 
the aim of this paper to present a clarification.  To avoid 
unnecessary complications, I begin by considering  the theory of two 
interacting scalar fields, one massive and one massless,
and compute the radiative corrections to the Casimir force
in  case of two plates separated by the distance $L$. 
Then  I generalize these results on the Casimir 
effect in QED\footnote{Note that 
the radiative corrections to the Casimir force are of no 
phenomenological significance. The interest in the problem 
is primarily related to such theoretical issue as the quantum 
field theory with non-trivial boundary conditions.}.

I will start with the detailed  description of the full calculation 
of the one-loop corrections  to the Casimir force. 
Similar  to \cite{bordag,kwong}, 
the massless fields satisfy certain  boundary 
conditions on the plates but no boundary conditions 
are imposed on the massive fields. 
Due to different quantization conditions on massive 
and massless fields, the calculation of 
radiative  corrections to the Casimir force 
is similar to the calculation  of  scattering processes 
{\it in  an external background field}
(whose exact form is determined  by the  boundary conditions on the massless
field) that facilitates virtual transitions from massless to massive 
fields. It follows from this 
that boundary-free  effective field theories do not completely 
account for the low-energy limit of the problem, since they are derived 
in the {\it absence} of a non-trivial external background.  This 
observation invalidates the results of \cite{kwong} as being derived
from the {\it wrong} effective field theory. Finally, I construct 
the effective field theory appropriate for the calculation of the 
Casimir force between two infinitely conducting parallel plates.

\section{The Casimir force}

Consider the theory of  two scalar fields $\phi$ and $\chi$ whose dynamics 
is governed by the Lagrangian:
\be
{\cal L} = \frac {1}{2} \left ( \partial _\mu \phi \right )^2 
+ \frac {1}{2} \left ( \partial_\mu \chi \right )^2 
- \frac {1}{2} M^2 \chi^2 - \frac {g}{2} \phi \chi^2.
\label{Lagr}
\ee
External conditions are set by 
two parallel plates located at the points $z=0$ and $z=L$. 
The Dirichlet boundary conditions are imposed on the 
field $\phi$, i.e. 
$\phi |_{z=0} = \phi |_{z =L} = 0$. 
We do not impose any boundary conditions 
on the massive field $\chi$.  The Green's function of the massless 
field $\phi$ is given by:
\be
G(t,\r_\perp,z,z') = \frac {2i}{L}\int \frac {d\omega}{2\pi} 
\frac {d^d k_\perp}{(2\pi)^d} 
e^{-i\omega t + i\k_\perp \r_\perp}\sum \limits_{n=1} 
\frac {\sin(\kappa n z) \sin (\kappa n z')}{ \omega^2 - \k_\perp^2 
- (\kappa n)^2 + i\delta}, 
\ee
with $\kappa = \pi/L$ and
$\k_\perp$ and $\r_\perp$ being momentum and 
position vectors that are parallel to the plates. I intend to 
use dimensional regularization for the calculations in what follows;
for this reason the transverse space is considered to be $d$-dimensional, 
with $d = 2-2\ep$ where $\ep$ is the regularization parameter.

The quantity of interest is the vacuum energy per unit 
transverse volume  between the two  plates. 
It can be obtained  from $T_{00}$ component of the 
stress-energy tensor (which is the Hamiltonian density of the system):
\be
T_{00} = \frac {1}{2} \left ( \partial_0 \phi \right )^2 
      + \frac {1}{2} \left (\vec {\partial} \phi \right )^2 
   + \frac {1}{2} \left ( \partial_0 \chi \right )^2 
      + \frac {1}{2} \left (\vec {\partial}  \chi \right )^2 
      + \frac {1}{2} M^2 \chi^2 
      + \frac {g}{2} \chi^2 \phi.
\ee
We then compute:
\be
E = \int \limits_{0}^{L} {\rm d}z~\langle 0 | T_{00}(z) | 0 \rangle.
\ee

Consider  first the leading order contribution. In momentum space 
it reads:
\be
E = \sum \limits_{n=1} \int \frac {d\omega}{2\pi} 
\frac {d^d k_\perp}{(2\pi)^d} 
\frac {i}{\omega^2 - \k_\perp^2 - (\kappa n)^2 + i \delta} 
\frac {\omega^2 + \k_\perp^2 + (\kappa n)^2}{2}.  
\ee
To compute $E$ in the above equation  we first 
integrate over $\omega$ by taking the residues 
(it is important to keep in mind that   scaleless integrals 
in dimensional regularization are set  to zero).
The result is:
\be
E = \frac {1}{2} \sum \limits_{n=1} \int  
\frac {d^d k_\perp}{(2\pi)^d} \sqrt{\k_\perp^2 + (\kappa n)^2}.
\ee
To proceed further, we rescale $|\k_\perp |\to |\k_\perp | \kappa n$; after 
that the integration over $k_\perp$ and the summation over $n$ factorize.
Finally, we obtain the well-known result:
\be
E = \frac {1}{2} \left ( \frac {\pi}{L} \right )^{d+1}
\zeta(-3+2\ep) \frac {\Gamma(-3/2+\ep)}{(4\pi)^{d/2}\Gamma(-1/2)}
 = - \frac {\pi^2}{1440~L^3}. 
\label{lo}
\ee

Note that,  in principle, 
we should subtract from the above result the  vacuum energy 
computed in the boundary-free field theory. 
However, 
it is easy  to see that for the massless field such a contribution vanishes 
(since it is given by a scaleless integral). For this reason 
Eq.(\ref{lo}) gives the complete result for the Casimir vacuum energy.
In general, we will understand the subtraction of  boundary-free
contributions as the result 
of adding position independent counterterms to the Hamiltonian
density\footnote{Because of the normal ordering prescription, 
these counterterms  are usually not included into boundary-free  field theory 
calculations.}. These counterterms are computed in boundary-free 
field theory  and 
they are constructed in such a way that
the vacuum energy density in the boundary-free case vanishes.

The Casimir force between the two plates is minus derivative of the 
vacuum energy with respect to the distance between the plates. One 
obtains:
\be
 f_0 = - \frac {\partial }{\partial L} E = -\frac {\pi^2}{480 L^4}.
\label{casforce}
\ee

\section{Radiative corrections}

We now turn to  the consideration of the radiative corrections
to the Casimir force. There are three diagrams that contribute;
they are  shown in Fig.1. Note that because of the form of the 
Lagrangian Eq.(\ref{Lagr}), the self-interaction of the field 
$\phi$ is of higher order in the coupling constant and for this
reason is not considered here.

We  begin with the diagram Fig.1a, 
that describes the  vacuum polarization of the massless field 
$\phi$ caused by  the massive field $\chi$.  
Since  massless and massive fields are quantized differently,
there is a peculiarity in  writing down expressions for Feynman diagrams
in that  one does not  expect the momentum conservation to hold for all  
interaction vertices. On the other hand, it is clear  how the 
perturbative expansion in   position  space is constructed  
and we can derive the momentum representation if 
we start from there. The expression for 
the vacuum energy in position space is:
\be
\delta E^{(1a)} \sim g^2 \int \limits_{0}^{L} {\rm d}z {\rm d}z_1 {\rm d}z_2
G(z,z_1) P(z_1-z_2)^2 G(z_2,z),
\label{e9}
\ee
where $P(z)$ is the propagator of the field $\chi$. 
The important point is that all the integrations range from $0$ to $L$ 
since this is the only  region where the Green's function $G$ 
has the non-vanishing  support. 

Because of the boundary conditions, the eigenfunctions 
of the $\chi$ field are not orthogonal to the eigenfunctions 
of the $\phi$ field; as a consequence 
we get momentum non-conservation in the  $\phi \chi^2$ 
interaction vertex. Switching to the momentum space in Eq.(\ref{e9}),
one finds the following  representation for $\delta E^{(1a)}$:
\be
\delta E^{(1a)} = -\frac {i}{2} \sum \limits_{n=1}
\int \frac {d\omega}{2\pi} 
\int \frac {{\rm d} \Delta}{(2\pi)} 
\frac {d^d k_\perp}{(2\pi)^d} 
\frac {( \omega^2 + \k_n^2)}{2} \frac {1}{(\omega^2 - \k_n^2+i \delta)^2}
{\cal F}_D(\kappa n, \Delta)
\Pi_M \left ( \omega,\k_\perp,\Delta \right),
\label{vp}
\ee
where
\be
{\cal F}_D(\kappa n, \Delta) = 
\frac {4 (\kappa n )^2}{L}
\frac {(1 - \cos (\kappa n L) \cos( \Delta L )  )}
{( \Delta^2 -(\kappa n)^2 + i\delta )^2 },~~~~~~~~~~~~~
\k_n = (\k_\perp,\kappa n),
\label{distr}
\ee
and $i\Pi_M$ is the vacuum polarization 
function\footnote{The $i \delta$ 
prescription introduced in the function ${\cal F}_D$ 
is for the integration convenience; it can be 
changed in an arbitrary way since the function is finite at 
$\Delta = \pm \kappa n$ because $\cos(\kappa L n) = (-1)^n$.}. 
It is apparent from Eq.(\ref{vp}) 
that there is a mismatch in the momentum that flows along 
the $\phi$ line and the 
momentum that goes into the vacuum polarization. The mismatch 
is described by the function ${\cal F}_D(\kappa n,\Delta)$; 
this function is peaked  at  $|\Delta| = \kappa n$, but 
it differs from zero for all other values of $\Delta$ and $\kappa n$.
 Note also that, for fixed boundary conditions,
the function ${\cal F}_D$  is universal and will appear 
in all the other diagrams we  consider.

To proceed further, it is convenient to use  the dispersion representation 
for the vacuum polarization function: 
\be
\Pi_M(q) = \frac {1}{\pi} \int \limits_{4M^2}^{\infty}
\frac {{\rm d}s~{\rm Im} \Pi_M(s)}{s - q^2 -i \delta},
\ee
where $q^2 = \omega^2 - \k_\perp^2 - \Delta^2$ and the imaginary 
part is defined as:
\be
{\rm Im} \Pi_M(s) = \frac {g^2}{(2\pi)^D} 
\frac {\pi^{7/2-\ep}}{2^{1-2\ep} \Gamma(3/2-\ep)} 
s^{-\ep} \left (1 -\frac {4M^2}{s} \right )^{1/2-\ep}.
\ee

We now see that the integration over $\Delta$ in Eq.(\ref{vp}) can 
be performed by using the residue theorem. There are two different 
contributions:  one from 
$\Delta = \pm \kappa n$ and another from 
$\Delta = \pm \sqrt{s - \omega^2 +\k_\perp^2}$. Let us consider these
two contributions separately. In view of the fact that the integral is 
symmetric with respect to the transformation $\Delta \to - \Delta$, 
we can substitute  $\cos (\Delta L) \to e^{i\Delta L}$ in the function 
${\cal F}_D$ and close the integration contour in the upper half plane.

Consider  first the contribution that comes from the second order 
residue at  $\Delta = -\kappa n + i \delta$. Since $\kappa n L = \pi n$, 
the only possibility to get a non-zero contribution is to differentiate 
$(1 - \cos(\kappa n L) e^{i \Delta L})$ in the 
function ${\cal F}_D$ with respect to $\Delta$. We  obtain:
\be
\delta E_{1}^{(1a)} = -\frac {i}{2}
\sum \limits_{n=1} \int \frac {d\omega}{2\pi} 
 \frac {d^d k_\perp}{(2\pi)^d} 
\frac {( \omega^2 + \k_n^2)}{2} \frac {1}{(\omega^2 - \k_n^2+i \delta)^2}
\Pi_M \left ( \omega,k_n \right ).
\label{vp1}
\ee
Note that this is exactly the contribution one would write down 
if momentum conservation in $\phi \chi^2$ vertex is  used blindly. 

To ensure that the field $\phi$ stays massless
one has to add a  mass counterterm to the above 
expression. Note, that since the subtraction is done at $q^2=0$, 
$\delta E_{1}^{(1a)}$ is the right place to add  the mass counterterm
contribution.  It is also convenient to perform the Wick 
rotation $\omega \to  i \omega$ at this step. 
We then obtain:
\be
\delta E_{1}^{(1a)} = \frac {1}{2}
\sum \limits_{n=1} \int \frac {d\omega}{2\pi} 
 \frac {d^d k_\perp}{(2\pi)^d} 
\frac {( \omega^2 - \k_n^2)}{2 k_n^2} 
\frac {1}{\pi} \int \limits_{4M^2}^{\infty}
\frac {{\rm d}s~{\rm Im} \Pi_M(s)}{s (s + k_n^2)},
\ee
where $k_n = (\omega,\k_n)$.
The simplest  way to proceed further is to do the partial 
fractioning to obtain:
\be
\frac {( \omega^2 - \k_n^2)}{2k_n^2} \frac {1}{(s+k_n^2)}
=\frac {\omega^2}{s} \frac {1}{k_n^2} 
- \frac {1}{2} 
\left (1 + \frac {2\omega^2}{s} \right) \frac {1}{k_n^2+s}.
\label{parfrac}
\ee
We now consider the two terms separately. The first one is:
\be
\delta E_{11}^{(1a)} = \frac {1}{2} 
 \sum \limits_{n=1} \int  \frac {d\omega}{2\pi} 
 \frac {d^d k_\perp}{(2\pi)^d}  
\frac {\omega^2}{k_n^2}  
\frac {1}{\pi} \int \limits_{4M^2}^{\infty}
\frac {{\rm d}s~{\rm Im} \Pi_M(s)}{s^2}.
\label{vp1a}
\ee
Taking into account that  scaleless integrals in dimensional 
regularization are set to  zero, one immediately recognizes 
that this contribution is  related to the wave function renormalization 
of the massless field and is therefore not relevant.

To deal with the second term  in the r.h.s. of Eq.(\ref{parfrac}), we first 
perform the sum over $n$ and neglect the terms that 
are exponentially suppressed $\le {\cal O}(e^{-2LM})$ (cf. the complete 
formula in Appendix):
\be
\sum \limits_{n=1}^{\infty}\frac {1}{\omega^2 + \k_\perp^2+ (\kappa n)^2+s}
= \frac {L}{2\sqrt{\omega^2 + \k_\perp^2 +s }} - 
\frac {1}{2 ( \omega^2 + \k_\perp^2 +s )} + {\cal O}(e^{-2LM}).
\label{ab}
\ee
We see that the middle term in the right hand side of Eq.(\ref{ab}) 
is $L$-independent  and hence does not contribute to the Casimir force.
On the other hand, the first term in the r.h.s. of Eq.(\ref{ab})
is nothing but the integral:
\be
L \int \frac {{\rm d}k_z}{(2\pi)} 
\frac {1}{  \omega^2 + \k_\perp^2+  k_z^2 +s} =
 \frac {L}{2\sqrt{\omega^2 + \k_\perp^2 +s }}, 
\ee
and hence its  contribution to the vacuum energy 
can be written as:
\be
\delta E_{12}^{(1a)} = - \frac {L}{4} 
\int \frac {d\omega}{2\pi} 
 \frac {d^{d+1} k}{(2\pi)^{d+1}}
 \frac {1}{\pi} \int \limits_{4M^2}^{\infty}
\frac {{\rm d}s~{\rm Im} \Pi_M(s)}{s  (\omega^2 + \k^2 +s)}
\left (1 + \frac {2\omega^2}{s} \right).
\label{resvp1}
\ee

A nice feature of this result is that it comes out 
being proportional to the distance between the two plates and so we can 
check that it will be  canceled by the boundary-free field theory 
counterterm for the vacuum energy density. 
The boundary-free vacuum energy  is given by the  expression:
\be
\delta E_{\rm free}^{(1a)} = -\frac {iL}{2} 
\int \frac {d\omega}{2\pi} 
 \frac {d^{d+1} k}{(2\pi)^{d+1}} 
\frac {( \omega^2 + \k^2)}{2} \frac {1}{(\omega^2 - \k^2+i \delta)^2}
\Pi_M \left ( \omega,\k \right ).
\ee
We again use (subtracted) dispersion relation for the vacuum polarization 
and proceed along the lines described above. We then find the 
result for $\delta E_{\rm free}^{(1a)}$ that exactly matches  
the result for  $\delta E_{12}^{(1a)}$ in Eq.(\ref{resvp1}). We therefore
conclude that, neglecting exponentially suppressed terms, 
the ``momentum conserving'' contribution to the Casimir 
energy $\delta E_{1}^{(1a)}$ 
vanishes upon  the subtraction of the boundary-free  
vacuum energy and appropriate renormalization of the wave function of the 
massless field.

Hence,  we have to concentrate on the second contribution to 
$\delta E^{(1a)}$ that comes from the ``momentum non-conserving''
part of the $\Delta$ integral in Eq.(\ref{vp}).  It is convenient 
to perform the Wick rotation $\omega \to i \omega$ and 
then integrate over $\Delta$ picking up the pole at 
$\Delta = i \sqrt{w^2 + \k_\perp^2 +s}$. We again neglect the exponentially
suppressed term $\le {\cal O}(e^{-2ML})$ coming from $\cos(\Delta L)$
and obtain:
\be
\delta E_{2}^{(1a)} = 
-\frac {1}{L} \sum \limits_{n=1} \int \frac {d\omega}{2\pi} 
 \frac {d^d k_\perp}{(2\pi)^d} 
\frac {( \omega^2 - \k_n^2)}{2(k_n^2)^2} 
\frac {1}{\pi} \int \limits_{4M^2}^{\infty}
\frac {{\rm d}s~{\rm Im} \Pi_M(s)}{\sqrt{s+w^2 + \k_\perp^2}} 
\frac {(\kappa n)^2}{(s+k_n^2)^2}.
\ee

To compute this integral, we combine the denominators using 
Feynman parameters and  carry out the integration over 
$\omega$. After appropriate rescaling, we integrate over $k_\perp$ and
introduce the  Mellin transform  to perform the sum
over $n$. The final result then reads (we do not 
display the  $L$-independent contribution):
\be
\delta E_{2}^{(1a)} = \frac {1}{24\pi^2 L} 
\int \limits_{4M^2}^{\infty} 
\frac {{\rm d}s~{\rm Im} \Pi_M(s)}{s}
\sum \limits_{n=1}^{\infty} 
(2n+3)\zeta(-2n-3)
\frac {\Gamma(n-1/2)}{\sqrt{\pi} \Gamma(n)}
\left ( \frac {\kappa}{\sqrt{s}}  \right )^{2n+3}. 
\label{ser1a}
\ee
We explicitly see that all these contributions are finite and 
depend in a non-analytic way on the mass of the heavy particle squared.
The first few terms read:
\be
\delta E_{2}^{(1a)} = -\frac {5}{12386304} \frac {\pi^4 \alpha}{M^3 L^6}
 +\frac {7}{150994944}\frac {\pi^6 \alpha}{M^5 L^8},
\label{res1a}
\ee
where  dimensionless coupling constant $\alpha = g^2/(4 \pi M^2)$ 
is introduced.

The series in Eq.(\ref{ser1a})  are asymptotic; the generating function
can be obtained  using the formulas from  Appendix:
\be
\delta E^{(1a)} = \frac {-1}{24 \pi^2 L}  
\int \limits_{4M^2}^{\infty} 
\frac {{\rm d}s~{\rm Im} \Pi_M(s)}{s}
\frac {1}{T(s) \pi } \int \limits_{0}^{\infty} 
\frac {{\rm d} u}{e^{u/T(s)}-1}
\frac {u^5 (5 + 4 u^2)}{(1+u^2)^{3/2}},
\label{fintem}
\ee
where $T(s) = \kappa /(2\pi \sqrt{s} ) $. The above equation completes 
our discussion of the  graph Fig.1a.

Let us  consider the 
graph shown in Fig.1b.  In the momentum representation  its 
contribution to the vacuum energy reads:
\be
\delta E^{(1b)} = -i \sum \limits_{n=1}
\int \frac {d\omega}{2\pi} 
\int \frac {{\rm d} \Delta}{(2\pi)} 
\frac {d^d k_\perp}{(2\pi)^d} 
\frac {1}{(\omega^2 - \k_n^2+i \delta)}
{\cal F}_D(\kappa n, \Delta)
V\left ( \omega,k_\perp,\Delta \right),
\label{vpb}
\ee
where ${\cal F}_D$ is defined in Eq.(\ref{distr}), 
the function $iV$ is given by 
\be
iV(q) = g^2 \int \frac {{\rm d}^D l}{(2\pi)^D} 
\frac {(2(Q l)^2 - l^2 + M^2)/2}{[l^2 - M^2+i\delta ]^2[(l+q)^2-M^2+i\delta]},
\ee
and $Q$ denotes the time-like vector $Q=(1,0,0,0)$.

We now give an  argument that no non-trivial momentum conserving contribution 
can appear from this graph. The argument goes as follows. Let us perform
the $\Delta$ integration by picking up the momentum conserving 
pole in $\Delta$. We get:
\be
\delta E^{(1b)} = -i \sum \limits_{n=1}
\int \frac {d\omega}{2\pi} 
\int \frac {{\rm d} \Delta}{(2\pi)} 
\frac {d^d k_\perp}{(2\pi)^d} 
\frac {1}{(\omega^2 - \k_n^2+i \delta)}
V\left (k_n \right).
\label{huh}
\ee
Let us now use the fact that the function $V(k)$ obeys dispersion relations 
with respect to  momentum transfer $k$. The dispersion 
relations are written in the subtracted form, to ensure that 
there is no correction to the interaction of the on-shell 
massless particle with  the stress-energy
tensor. We then have:
\be
V(k_n) = k_n^2  \left [ \frac {1}{\pi} \int \limits_{4M^2}^{\infty}
\frac {{\rm d}s~{\rm Im} V_1(s)}{s(s - k_n^2 -i \delta)} +
(Qk_n)^2 \frac {1}{\pi} \int \limits_{4M^2}^{\infty}
\frac {{\rm d}s~{\rm Im} V_2(s)}{s^2(s - k_n^2 -i \delta)} \right ],
\ee
where the imaginary parts are given by:
\ba
&& {\rm Im} V_1(s) = \frac {-g^2}{(2\pi)^D} 
\frac {\pi^{7/2-\ep}}{2^{3-2\ep} \Gamma(3/2-\ep)} 
s^{-\ep} \left (1 -\frac {4M^2}{s} \right )^{1/2-\ep}, \\
&&
{\rm Im} V_2(s) = \frac {g^2}{(2\pi)^D} 
\frac {\pi^{7/2-\ep}}{2^{2-2\ep} \Gamma(3/2-\ep)} 
\left ( \ep - \frac {2M^2}{s}  \right ) 
s^{-\ep} \left (1 -\frac {4M^2}{s} \right )^{-1/2-\ep}.
\ea

If we substitute this expression to Eq.(\ref{huh}), we  
cancel the massless particle propagator; the only other ``propagator'' 
left is  $1/(s-k_n^2)$. Similar to the case of the vacuum polarization 
graph discussed above, the sum over $n$ can be performed; disregarding  
the exponentially suppressed contribution and also the contribution 
which is $L$ independent, this sum appears equivalent to the 
integral over $k_z$. It is then easy to see  that this contribution
exactly matches the boundary-free vacuum energy and  gets 
subtracted completely.

Therefore, the only non-trivial contribution to the vacuum energy from 
Fig.1b  is again related to 
the momentum configuration where $\Delta$ is of the order of the 
mass of the heavy particle.  To compute this contribution we proceed
in a way similar to what has been done for 
the vacuum polarization graph Fig.1a and 
obtain:
\ba
\delta E^{(1b)} && = \frac {1}{2\pi^2 L}
\int \limits_{4M^2}^{\infty} 
\frac {{\rm d}s~{\rm Im} V_1(s)}{s}
\sum \limits_{n=1}^{\infty} \zeta(-2n-1)
\frac {\Gamma(n-1/2)}{\sqrt{\pi} \Gamma(n)} 
\left ( \frac {\kappa}{\sqrt{s}} \right )^{2n+1}
\nonumber \\
&& + \frac {1}{3\pi^2 L}
\int \limits_{4M^2}^{\infty} 
\frac {{\rm d}s~{\rm Im} V_2(s)}{s}
\sum \limits_{n=1}^{\infty}\zeta(-2n-1)
\frac {\Gamma(n-1/2)}{\sqrt{\pi} \Gamma(n)} \frac {(n-1)}{(2n-3)} 
\left ( \frac {\kappa}{\sqrt{s}} \right )^{2n+1}.
\ea

Again, the series are asymptotic; the generating  function 
for these series can  be easily established in 
a manner similar to what has lead to Eq.(\ref{fintem}). The first few terms of 
$\delta E^{(1b)}$ read:
\be
\delta E^{(1b)} = -\frac {1}{245760} \frac {\pi^2 \alpha}{L^4 M}
 +\frac {13}{24772608} \frac {\pi^4 \alpha}{L^6 M^3}.
\label{res1b}
\ee

The final contribution one has to consider comes from the diagram
shown in Fig.1c and reads:
\be
\delta E^{(1c)} = -\frac {i}{2} \sum \limits_{n=1}
\int \frac {d\omega}{2\pi} 
\int \frac {{\rm d} \Delta}{(2\pi)} 
\frac {d^d k_\perp}{(2\pi)^d} 
 \frac {1}{(\omega^2 - \k_n^2+i \delta)}
{\cal F}_D(\kappa n, \Delta)
\Pi_M \left ( \omega,k_\perp,\Delta \right).
\label{1c}
\ee
This contribution is easily analyzed following the discussion 
of the vacuum polarization graph Fig.1a at the beginning of this 
Section. Again, the only $L$-dependent contribution that 
does not get canceled against the corresponding boundary-free field 
theory counterterm is related to large values of $\Delta$. 
The full result is:
\be
\delta E^{(1c)} = \frac {1}{4 \pi^2 L } 
\int \limits_{4M^2}^{\infty} 
\frac {{\rm d}s~{\rm Im} \Pi_M(s)}{s}
\sum \limits_{n=1}^{\infty} \zeta(-2n-1)
\frac {\Gamma(n-1/2)}{\sqrt{\pi} \Gamma(n)}
\left ( \frac {\kappa}{\sqrt{s}}  \right )^{2n+1}, 
\ee
and the first few terms read:
\be
\delta E^{(1c)} = 
 \frac {1}{122880} \frac {\pi^2 \alpha}{L^4 M}
 -\frac {1}{4128768} \frac {\pi^4 \alpha}{L^6 M^3}.
\label{res1c}
\ee

The sum of the above results Eqs.(\ref{res1a},\ref{res1b},\ref{res1c}) 
gives the correction to the vacuum 
energy:
\be
\delta E^{(1)} = \frac {1}{245760} \frac {\pi^2 \alpha}{L^4 M} 
+{\cal O}(\alpha L^{-6}M^{-3}).
\ee
The relative correction to the Casimir force then becomes:
\be
\frac {\delta f}{f_0} =  -\frac {\alpha}{128 LM},
\label{casfsc}
\ee
where $f_0$ is the leading order result Eq.(\ref{casforce}).
Let me emphasize that this contribution is non-analytic in the 
mass of the heavy particle squared
and in this respect is similar to the result 
reported in \cite{bordag}\footnote{To avoid confusion, Eq.(\ref{casfsc})
is derived in the context of the scalar field theory, 
whereas the calculation of Ref.\cite{bordag} is done for QED.}.

\section{Effective field theory description}

My main motivation for considering this problem was the 
desire to clarify the apparent  discrepancy between the 
full  and the effective field theory 
calculations of the QED radiative corrections to the Casimir 
energy in case of two infinitely conducting parallel plates. 
Having performed an explicit full
calculation in the previous Section, we can establish what 
went wrong with the effective field theory arguments as 
applied in Ref.\cite{kwong} and the way to modify them 
to make the effective field theory approach work. 

Our key (and rather simple) observation is the fact that the 
momentum in the heavy-light  vertex is not conserved, as follows
from  different quantization conditions on  massless and 
massive  fields.
In this situation one can think about heavy-light transitions 
as being induced by external potential. To compute the total energy,
one has to integrate over the momentum transfer $\Delta$ from the external 
potential to the quantum fields.  This momentum transfer can be 
both small $\Delta \sim L^{-1}$ and large $\Delta \sim M$. 

When the momentum transfer is small $\Delta \sim L^{-1}$, one can 
use  boundary-free effective  field theory, produced by integrating
out heavy particles, in the calculation of the Casimir 
force. From the diagrammatic viewpoint the ``hard'' (with $k \sim M$) 
subgraph in this case   is the loop made up of only massive lines
(see Fig.2 ) and is clearly the same as in case of boundary-free 
field theory. This momentum configuration does produce some 
effective Lagrangian; however, as we have seen in the previous  
Section it has no effect on the Casimir force if the 
usual renormalization program is adopted. 

Let me note in passing that, in principle, 
there is another contribution coming from the 
momentum conserving piece of the interaction. It corresponds
to the situation when the momentum of the massless particle
 becomes 
large  ( in other words, it corresponds to  large values 
of $n$ in the sum).  As I have shown in the previous Section, 
if one neglects the exponentially suppressed terms and the 
terms that are $L$-independent, the contribution 
coming from this momentum region matches exactly the boundary free 
contribution  to the vacuum energy. The trick there was to use the dispersion 
relations with the cut starting from $4M^2$ and it is not completely clear 
to me how this particular contribution can be argued 
away in a more general case. 

The  configuration with large momentum transfer $\Delta \sim M$ 
does not  have an analogy in the boundary-free case; 
it is peculiar to the existence 
of the non-trivial potential.  However, apart from the 
very existence of this contribution, there is nothing special 
about it and one can treat it  along  conventional 
lines of the effective field theories; the change
is in the  subgraphs that determine  effective field 
theory operators with their Wilson coefficients.

Provided that dimensional regularization is used for the calculations, 
it is easy to give a general prescription for computing this contribution 
in a simple way. Following the general pattern of  asymptotic expansions
of Feynman diagrams, one has to identify small and large quantities 
for a given momentum configuration and then Taylor expand in all the small 
quantities. For the case $\Delta \sim M$, the small quantities are the 
momentum components of the massless field and so one has to Taylor expand 
in these variables and integrate over the large components of the loop momenta.
One then explicitly generates a set of local operators with  
corresponding Wilson coefficients. The vacuum polarization (Fig.1a) induces 
the following contribution to the effective Lagrangian: 
\be
\delta {\cal L}  = - \frac {\alpha }{512  L M}
 \phi \left [ \partial _z^2 \right] \phi 
 +  \frac {\alpha}{8192 M^3 L} \phi \left  [ \partial_z^4
 + 5 \partial ^2 \partial _z ^2 \right ] \phi. 
\label{efflagr}
\ee
The other two graphs (Fig1.b,c) represent the renormalization 
of the $T_{00}$  component of the stress-energy  tensor.  
The corresponding  shift in $T_{00}$ can be either 
computed directly or obtained from $\delta {\cal L}$ in the standard way.

Let me repeat that the complete low energy effective field 
theory for the field $\phi$ also includes  terms in addition 
to those shown in $\delta {\cal L}$ . 
These terms are $L$-independent, analytic in the mass of the heavy particle 
squared and can be derived by working with the boundary-free field theory. 
As I have shown in the previous Section, 
those terms do not produce any non-trivial $L$-dependence and 
are therefore not relevant for the Casimir force. 

Finally, let me show that the effective Lagrangian $\delta {\cal L}$ 
does indeed produce the required correction to the Casimir force.
To that end, we  consider two sources of the correction: the
first one is the correction to the Green's function of the $\phi$ field
induced by the effective  Lagrangian $\delta {\cal L}$ 
Eq.(\ref{efflagr}); the second one is the  correction to 
the stress-energy tensor. 

Consider  the change in the vacuum energy caused by 
the correction to the Green's function of the field $\phi$. 
We can write it in the following way:
\be
\delta E^{(1a)}_{\rm eff} = -
\sum \limits_{n=1} \int \frac {d\omega}{2\pi} 
\frac {d^d k_\perp}{(2\pi)^d} 
\frac {\omega^2 + \k_n^2}{2 (k_n^2)^2}
\left ( \frac {i\alpha}{256LM} (\kappa n)^2 
+ \frac {i \alpha}{4096M^3 L} \left [(\kappa n)^4 + 5 (\kappa n)^2 k_n^2  
\right ]\right ).
\ee
Further calculation is easy if one performs the Wick rotation and integrates
over $\omega$ and $\k_\perp$. The leading ${\cal O}(M^{-1})$ 
contribution in the above equation then vanishes and the result reads:
\be
E^{(1a)}_{\rm eff} = -\frac {5}{12386304} \frac {\pi^4 \alpha}{M^3 L^6},
\ee
in  agreement with the first term in Eq.(\ref{res1a}).

The second contribution occurs because of  the change in the stress-energy 
tensor (it corresponds to graphs  Fig.1b,c) 
which can be easily derived from $\delta {\cal L}$:
\be
\delta T_{00} = 
\frac {\alpha }{512  L M}
 \phi \left [ \partial _z^2 \right] \phi 
 +  \frac {\alpha}{8192 M^3 L} \phi \left  [- \partial_z^4
 + 5 \left ( \partial_0 ^2 + \vec \partial ^2  \right ) 
\partial _z ^2 \right ] \phi. 
\ee
Such a change in $T_{00}$ produces the following shift in the vacuum 
energy:
\be
\delta E^{(1b)}_{\rm eff} = 
\sum \limits_{n=1} \int \frac {d\omega}{2\pi} 
\frac {d^d k_\perp}{(2\pi)^d} 
\frac {i}{k_n^2} \left ( 
-\frac {\alpha}{512 LM} (\kappa n)^2 + \frac {\alpha}{8192 M^3 L} 
 \left [ 5 (\omega^2 + \k_n^2) (\kappa n)^2 - (\kappa n)^4 \right ] 
\right ).
\ee
The remaining calculation is a simple matter and one obtains:
\be
\delta E^{(1b)}_{\rm eff} = \frac {1}{245760} \frac {\pi^2 \alpha}{L^4 M}
+ \frac {1}{3538944} \frac {\pi^2 \alpha}{M^3 L^6}.
\ee
Again, we find an agreement between this result and the sum of 
Eq.(\ref{res1b}) and Eq.(\ref{res1c}). We therefore see that it 
is possible to derive an effective field theory which produces 
corrections to the Casimir force that are in  agreement with the 
full calculation.

\section{Radiative corrections in QED}

With the understanding gained by considering the scalar field 
theory example above, it is now a relatively simple matter 
to compute the radiative corrections to the Casimir force in 
QED. The only difference with the scalar field theory is 
that the quantization procedure is more tedious. 
For this reason I am going to spell it out in some detail.

We consider two infinitely conducting plates and impose 
standard boundary conditions on  electric and magnetic fields:
\be
\vec E_{\perp} = 0,~~~~~~~\vec B_{z} = 0.
\label{bcqed}
\ee
The electron field is allowed to propagate freely. This 
set of boundary conditions is the same as in Refs.\cite{bordag,kwong}. 

I quantize the theory in the Coulomb gauge. For  transverse degrees
of freedom, the required  formalism has been developed in \cite{rl}
where explicit expressions for the vector potentials satisfying 
boundary conditions  Eq.(\ref{bcqed}) have been given. Using these 
results, we obtain the following propagator for the transverse 
photon field:
\be
\langle  {\bf A}^i(t,\r_\perp,z) {\bf A}^j(0,{\bf 0},z') \rangle  = 
\frac {2i}{L}\sum \limits_{n}^{} 
\int \frac {{\rm d}^dk_\perp}{(2\pi)^d} \frac {{\rm d}\omega}{2\pi}
e^{i\k_\perp \r_\perp -i \omega t }
\left ( \delta^{ij} - \frac {\k_n^i \k_n^j}{\k_n^2} \right ) 
\frac {\vec X_i(\kappa n z) \vec X_j^{*}(\kappa n z')}
{\omega^2 -\k_n^2+i\delta},
\label{phprop}
\ee
where the vector  $\vec X$ is given by:
\be
\vec X = [ i \sin( \kappa n z), i \sin (\kappa n z), \cos (\kappa n z)]. 
\ee
Note that the normalization $2/L$ shown in Eq.(\ref{phprop}) 
only applies to modes with $n \ge 1$. However, for the $z$-component 
of the vector potential, there is a zero mode with $n=0$; its normalization 
factor is two times smaller. The zero mode will not 
be of any concern for us in what follows and for this reason I use
a (slightly misleading) representation for the photon propagator 
as in Eq.(\ref{phprop}). 

Let me now derive the leading order result for the Casimir force in 
QED. The photon part of the $
T_{00}$ component of the stress-energy tensor is:
\be
T_{00} = \frac {1}{2} \left [ \left (\partial_0 \vec A \right )^2 
+ \sum \limits_{i=1}^{3}
\left ( \partial _i \vec A \right )^2 \right ], 
\ee
and is therefore diagonal with respect to the components of the vector 
potential. This implies that only  average values of $\sin^2(\kappa n z)$
and $\cos^2(\kappa n z)$ will enter the formula for the Casimir energy and 
since these two averages are the same,  the result reads: 
\be
E^{(0)}_{\rm QED} = 
\sum \limits_{n=1} \int \frac {d\omega}{2\pi} 
\frac {d^d k_\perp}{(2\pi)^d} 
 \frac {\omega^2 + \k_n^2}{2} \delta_{ij} 
 \left ( \delta^{ij} - \frac {\k_n^i \k_n^j}{\k_n^2} \right ) 
\frac {i }{\omega^2 - \k_n^2 + i \delta }.
\ee
Performing the sum over polarizations, we observe that $E^{(0)}_{\rm QED}$
is two times  larger than the corresponding result for the 
scalar field in Eq.(\ref{lo}) and we obtain:
\be
E^{(0)}_{\rm QED} = - \frac {\pi^2}{720 L^3}.
\ee

Let us now turn to radiative corrections to this result.
As we have seen from the scalar field theory example, 
the important thing is the expression for  external potential.
Since this potential depends on the 
boundary conditions and they are different for different components 
of the vector potential, the potential is  sensitive to  photon 
polarization.  

To show the  kind of expressions  one gets, I consider
the vacuum polarization diagram Fig.1a and the corresponding 
correction to the vacuum energy:
\ba
E_{\rm QED}^{(1a)} = &&  
\sum \limits_{n=1} \int \frac {d\omega}{2\pi} 
\frac {d^d k_\perp}{(2\pi)^d} 
{\cal F}(\kappa n, \Delta )  
 \frac {\omega^2 + \k_n^2}{2} 
\delta_{ij} 
 D_{ii_1}(k_n)W_{i_1i_2}(\kappa n, \Delta)  
\Pi_{i_2 j_2}\left (\omega,\k_\perp,\Delta \right ) 
\nonumber \\
&& ~~~~~~~~~~~~~~~~~~~~~~~~\times
W_{j_2 j_1}(\kappa n, \Delta)  D_{j_1 j}(k_n),
\ea
where $\Pi_{i_2 j_2}$ is the vacuum polarization function,
$D_{ij}(k) = i ( \delta_{ij} - \k_i \k_j /\k^2 )/k^2$ is the 
usual photon propagator in the Coulomb gauge,
${\cal F}$ is defined be the equation:
\be
{\cal F} =  \frac {4}{L}
\frac {( 1-\cos(\kappa n L ) \cos(\Delta L) )}{[ \Delta^2 - (\kappa n)^2 ]^2},
\ee
and the matrix $W$ reads:
\be
W_{ij} = \kappa n  \left ( \delta_{ij} 
- {\vec n}_i {\vec n}_j \right ) +\Delta {\vec n}_i {\vec n}_j,
\ee
with  $\vec n$ being  the  unity vector in the $z$ direction. 

We see that the only significant change in the structure of the 
potential, as compared to the  scalar case, is that now it changes
the ``polarization'' of the photon in addition to the change in its 
momentum.  Although this is an important modification, it does not 
affect our discussion of the  relevant momentum regions in the 
previous Section. Therefore, the only momentum region that can give
a non-zero contribution is the the momentum configuration where $\Delta$ 
is of the order of the electron mass  while  
the photon momentum remains small. We can therefore proceed with the 
calculation of the effective Lagrangian in exactly the same way
as in the previous Section, i.e. we consider the vacuum polarization 
diagram and integrate out ``hard subgraphs'' that include the lines 
of the external ``potential''. We then 
obtain the effective Lagrangian accurate to ${\cal O}(1/m)$:
\be
\delta {\cal L}_{QED} 
= \frac {3 \alpha}{32 L m } \left ( - {\vec E}_z^2 + \vec B_\perp ^2 
 \right ).
\label{qedeff}
\ee

Again, there are two sources of radiative corrections caused by this 
effective Lagrangian; the first one is the correction to the Green's
function of the photon field and the second is the modification of the 
energy momentum tensor. The calculation of these corrections 
is now a simple  matter and we easily reproduce the result 
of \cite{bordag}:
\be
\delta E^{(0)}_{QED} = \frac {\alpha \pi^2 }{2560 m L^4}. 
\ee
The corresponding correction to the Casimir force is then:
\be
\frac {\delta f_0}{f_0} = -\frac {3}{8} \frac {\alpha}{Lm}. 
\ee

\section{Conclusion} 

In this paper the radiative corrections to the Casimir force 
between two parallel plates are studied in both scalar
field theory with one massive and one massless field and  in QED. 
It is  shown  that, due to different quantization conditions for massive 
and massless fields, it is not possible to use   
effective  field theories derived in the boundary-free case 
to get corrections  to the Casimir force. 

The picture that emerges from these considerations  
can be visualized by saying that the transition 
from massless to massive fields can not happen directly, but only 
through the interaction with the external potential. The expression 
for this potential is derived from the overlap integral 
of the eigenfunctions  of heavy and light fields and is therefore 
boundary conditions dependent.
Since the momentum  transfer from the potential to  quantum fields 
can be both large  and small, one has to take this into account 
while constructing effective field theory for the calculation 
of the Casimir force by integrating out heavy degrees of freedom.

As can be expected from general considerations, the operators in the effective 
Lagrangian do not respect Lorentz invariance since the presence 
of plates breaks translation invariance in the $z$-direction. 
For the case of two interacting scalar fields, the effective Lagrangian 
is shown in  Eq.(\ref{efflagr}). One sees explicitly that it is 
given by power series in $[\partial _z^2/M^2]$. The same can be seen
from the effective Lagrangian for the QED Casimir effect given in 
Eq.(\ref{qedeff}).

Another remark concerns the dependence of the result on the boundary
conditions. Consider the scalar field theory. In that case, 
all the results in the present paper correspond to 
Dirichlet boundary conditions for  the massless field
and one may wonder what will change if other boundary conditions 
are used. Let us choose  von Neumann conditions
$\partial_z \phi |_{z=0} = \partial_z \phi |_{z=L} = 0$. In this situation 
the Green's function of the field $\phi$ changes.  In spite of that, 
the leading order  Casimir force remains the same. The radiative corrections 
to the Casimir force, however, do change; this happens
because the function ${\cal F}_D(\kappa n, \Delta)$ depends on the boundary 
conditions. Such a dependence is not very surprising  
since large values of $\Delta$, that are responsible for the radiative 
corrections,  correspond to the values  $z \sim 0 $ and 
$z \sim L$ and it is clear that the behavior of the field $\phi$ close 
to the plates is sensitive to the boundary conditions.  
In this  respect the effective field theories for the Casimir force may be 
not that ``effective'' since the results of the present  paper show
that  it is not possible to first 
integrate out  heavy degrees of freedom and impose boundary conditions 
on remaining light degrees of freedom later on.

My final comment is on the relation of the results obtained 
in this paper and the calculation of radiative corrections to 
photon energy density at small temperature $T$. In fact, the 
calculation in Ref.\cite{kwong} has been first performed 
for the finite temperature and later argued to be equivalent 
to the calculation of radiative corrections to the Casimir 
force once the substitution $T \to 1/(2L)$ is made. As I have 
shown above, the results of Ref.\cite{kwong} related to  radiative 
corrections to the Casimir force are based on the wrong effective field 
theory and for this reason are not valid. The natural question is
then whether or not the finite temperature results of \cite{kwong}
are correct. To answer this question we have to recall that the formulation 
of the finite temperature field theories based on functional integral 
with certain boundary conditions on bosonic and fermionic degrees of 
freedom is just a (powerful) mathematical trick. However, this
trick is not necessary. To compute the energy density of the photon 
field, one can imagine  constructing (without any restrictions on space-time)
an effective Hamiltonian first  and  
taking thermal average over the photon states later.
It is crucial that no reference to boundary conditions 
appears in the process of constructing effective Hamiltonian in this 
case  and therefore heavy degrees of freedom  can be 
integrated out in the boundary-free manner. It is also physically sound 
to take thermal averages over the photon states or, 
equivalently,  impose  periodic boundary conditions  on the 
photon field in the functional integral 
{\it after}  heavy particles have been integrated out\footnote{Note that 
in this case the photons and the electrons are not in thermal equilibrium.}. 
This is exactly the procedure that has been followed in \cite{kwong}
and the above  comments show that it is meaningful in case of the finite 
temperature.  It is, however, not at all meaningful for the Casimir 
effect  where first  boundary conditions should be imposed and 
only after that should the heavy particles  be integrated out. In my opinion, 
the main result of this paper is a clear demonstration of the fact that 
these two procedures do not commute.

\section{Acknowledgments}
This work was supported in part by DOE under grant number
DE-AC03-76SF00515. I am grateful to Lance Dixon and Marvin 
Weinstein for many useful conversations. 

\section*{Appendix}

Some formulas used in the derivation.

\be
I_0(a,b) = \int \frac {d^d p}{(2\pi)^d}
\lb \fr{1}{p^2+1} \rb^a \lb \fr{1}{p^2}\rb^b
 = (4 \pi)^{-d/2}
 \frac {\Gamma(a+b-d/2)\Gamma(d/2-b)}{\Gamma(a)\Gamma(d/2)},
\ee

\be
\zeta(-2n-3) = \frac {(-1)^n \zeta(2n+4) (2n+3)!}{\pi^{2n+4} 2^{2n+3}},
~~~~~~~~~~~~~~~
\int \limits_{0}^{1} \frac {{\rm d}x~~\ln^n x}{1-x} = 
(-1)^n n! \zeta(n+1).
\ee

\be
\Gamma\left ( z+\frac {1}{2} \right ) = \frac {\sqrt{\pi}}{2^{2z-1}}
\frac {\Gamma(2z)}{\Gamma(z)},
~~~~~~~~~~~~~~~~
\sum \limits_{n=1}^{\infty} \frac {1}{x^2 + \pi^2 n^2} = \frac {1}{2x} \left ( \frac {2}{e^{2x}-1}
+1 - \frac {1}{x} \right ).
\ee

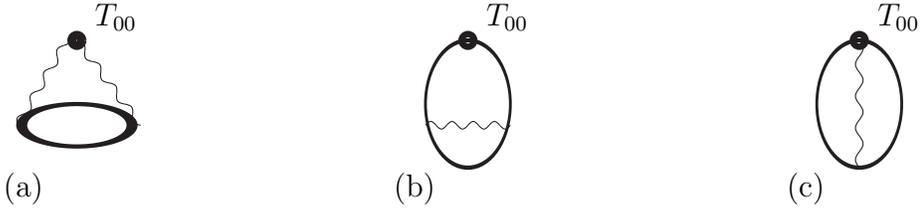
\begin{figure}%[h]
\begin{center}
\hfill
\begin{picture}(120,60)(0,0)
  \SetScale{.8}
\SetWidth{3}
 \Oval(50,60)(3,3)(0)
 \Oval(50,60)(1,1)(0)
 \SetWidth{2}
 \Oval(50,20)(10,26)(0)
\SetWidth{0.5}
  \Photon(22,20)(47,63){1.8}{4}
  \Photon(80,20)(50,60){1.8}{4}
 \Text(55,62)[t]{$T_{00}$}
 \Text(20,-2)[t]{(a)}
\end{picture}
\hfill
\begin{picture}(120,60)(0,0)
 \SetScale{.8}
\SetWidth{3}
 \Oval(50,60)(3,3)(0)
 \SetWidth{2}
 \Oval(50,30)(30,20)(0)
\SetWidth{0.5}
  \Photon(30,20)(70,20){1.8}{4}
 \Text(55,62)[t]{$T_{00}$}
\Text(20,-2)[t]{(b)}
\end{picture}
\hfill
\begin{picture}(120,60)(0,0)
 \SetScale{.8}
\SetWidth{3}
 \Oval(50,60)(3,3)(0)
 \SetWidth{2}
 \Oval(50,30)(30,20)(0)
\SetWidth{0.5}
  \Photon(50,0)(50,60){1.8}{4}
 \Text(55,62)[t]{$T_{00}$}
 \Text(20,-2)[t]{(c)}
\end{picture}
\hfill\null\\
\vglue 18pt
\end{center}
\caption{One loop  corrections to the vacuum energy. The wavy and bold 
solid lines
represent the massless and the massive particles, respectively.
Insertions of the stress-energy tensor are shown explicitly.}
\label{fig1}
\end{figure}

\begin{figure}[h]
\begin{center}
\hfill
\begin{picture}(120,60)(0,0)
  \SetScale{.8}
\SetWidth{3}
 \Oval(50,60)(3,3)(0)
 \Oval(50,60)(1,1)(0)
 \SetWidth{2}
 \Oval(50,20)(10,26)(0)
\SetWidth{0.5}
  \Photon(22,20)(47,63){1.8}{4}
  \Photon(80,20)(50,60){1.8}{4}
 \Line(22,20)(50,-5)
 \Line(76,20)(50,-5)
\SetWidth{1}
 \Line(50,-8)(50,-2)
 \Line(47,-5)(53,-5)
\SetWidth{0.5}
 \Text(55,62)[t]{$T_{00}$}
 \Text(20,-10)[t]{}
\end{picture}
\hfill\null\\
\vglue 18pt
\end{center}
\caption{One loop  correction to the vacuum energy with ``external 
potential'' lines shown explicitly. Momentum configuration that 
produces non-trivial Wilson coefficients corresponds to the large 
$\sim M$ momenta flowing through the massive {\it and} external potential 
lines.}
\label{fig2}
\end{figure}
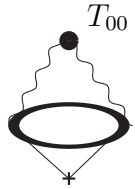

\end{document}